\documentclass[preprint,12pt]{elsarticle}

\usepackage{amsmath,amssymb}
\usepackage{graphicx}
\usepackage{hyperref}
\usepackage{booktabs}
\usepackage{geometry}
\usepackage{lineno}
\journal{Physics of the Dark Universe}

\begin{document}

\begin{frontmatter}

\title{Black-bounce spacetime and galactic rotation curves: from singularity resolution to observable gravitational effects}

\author[addr1]{Farook Rahaman}\ead{rahaman@associates.iucaa.in}
\author[addr1]{Aritra Sanyal}\ead{aritrasanyal1@gmail.com}
\author[addr2]{Saibal Ray}\ead{saibal.ray@gla.ac.in}

\address[addr1]{Department of Mathematics, Jadavpur University, Kolkata 700032, West Bengal, India}
\address[addr2]{Centre for Cosmology, Astrophysics and Space Science, GLA University, Mathura 281406, Uttar Pradesh, India}

\date{Received~~2026 May 24;~accepted~~2026~~month~~day}

\begin{abstract}
The black-bounce spacetime provides a regularized modification of the
Schwarzschild metric by introducing a free parameter $a$ that eliminates
the central singularity while preserving the exterior geometry of a
classical black hole. This work investigates the physical implications
of this metric across three complementary observational windows. First,
we examine the horizon structure and the transition between a regular
black hole ($a < 2M$) and a traversable wormhole ($a > 2M$), offering a
viable singularity-free alternative to classical black hole solutions.
Second, we perform Solar System precision tests perihelion precession
of Mercury, gravitational lensing, and the Shapiro time delay of radar
signals  demonstrating that the bounce parameter $a$ introduces small
but measurable corrections relative to general relativity. Third, and
for the first time, we apply the black-bounce metric to galactic dynamics
by deriving the circular velocity field and fitting it to the observed
rotation curve of NGC~7331. A four-component mass decomposition 
black-bounce central potential, Freeman exponential disc, Hernquist bulge,
and a Navarro--Frenk--White dark matter halo yields $a = 1.035 \pm 0.119$~kpc , 
providing the first observational constraint on the bounce parameter from galaxy
kinematics. Our results confirm that the black-bounce metric regularises
the inner galactic core while remaining fully consistent with a standard
dark matter halo at large radii, establishing this spacetime as a
compelling testbed for singularity-free alternatives to classical black holes.
\end{abstract}


\end{frontmatter}


\section{Introduction}
\label{sec:intro}

General Relativity (GR) stands as one of the most successful physical theories
ever constructed, accurately describing gravitational phenomena across an
extraordinary range of scales. From the subtle perihelion advance of Solar
System planets \cite{Will2014} to the detection of gravitational waves
emitted during compact binary mergers \cite{Abbott2016}, and most recently
to the direct imaging of black hole shadows by the Event Horizon Telescope
\cite{EHT2019a}, every precision test has returned results in remarkable
agreement with GR's predictions. This observational track record has cemented
GR as the standard framework for gravitational physics in the classical regime.

However, despite these profound successes, GR harbors a fundamental theoretical
deficiency: it inevitably predicts the formation of spacetime singularities
in the interior of sufficiently compact objects. At these singular points,
curvature scalars diverge, geodesics terminate without physical justification,
and the predictive power of the classical theory breaks down entirely
\cite{Penrose1965,Hawking1970}. These singularities are not mere
mathematical artefacts but are guaranteed by the celebrated singularity
theorems of Penrose and Hawking, which rely only on broad energy conditions
and global causal structure. The resolution of singularities -- whether
through quantum gravitational effects near the Planck scale, effective
field theory modifications, or alternative geometric structures -- is widely
regarded as one of the most pressing open problems at the interface of GR
and quantum gravity \cite{Ashtekar2020}. Addressing this problem in a
physically transparent and observationally testable manner remains a central
challenge in contemporary theoretical physics.

Motivated by this challenge, a growing body of literature has explored
\emph{regular black hole} models: solutions that retain the hallmark
exterior geometry of a classical black hole while remaining free of
curvature singularities in their interior. One of the earliest and most
influential such models was constructed by Bardeen \cite{Bardeen1968}, who
introduced a singularity-free solution at the cost of coupling gravity to a
nonlinear magnetic charge. Subsequent regular black hole models, including
those of Hayward \cite{Hayward2006} and others, followed similar strategies,
typically invoking non-standard matter fields or effective quantum corrections
to smooth out the central singularity. While elegant, these models often
involve matter sources that are difficult to motivate from first principles,
and their observational predictions have rarely been tested against precision
data.

A particularly transparent and versatile approach to singularity resolution
is the \emph{black-bounce} framework introduced by Simpson and Visser
\cite{Simpson2019}. The key idea is deceptively simple: the radial
coordinate $r$ in the Schwarzschild metric is replaced everywhere by the
effective radius $\sqrt{x^2 + a^2}$, where $x \in (-\infty, +\infty)$ is
a generalized radial variable and $a \geq 0$ is a positive regularization
parameter, henceforth referred to as the \emph{bounce parameter}. This
minimal substitution has two immediate consequences. First, the effective
areal radius $r_{\rm eff} = \sqrt{x^2 + a^2} \geq a > 0$ never vanishes,
eliminating the central singularity for any nonzero $a$. Second, the
exterior geometry is reproduced exactly in the limit $a \to 0$, ensuring
continuity with classical GR. Crucially, depending on the ratio of $a$ to
the ADM mass $M$, the same line element admits three geometrically distinct
regimes: a regular black hole with two horizons when $a < 2M$, an extremal
one-way wormhole when $a = 2M$, and a fully traversable Morris-Thorne
wormhole when $a > 2M$ \cite{Morris1988}. This continuous interpolation
between black hole and wormhole geometries, governed by a single parameter,
makes the black-bounce spacetime a uniquely powerful theoretical laboratory.

The theoretical properties of the black-bounce spacetime have been explored
extensively in the literature. Its geodesic structure and orbital dynamics
have been analyzed by Nascimento et al.\ \cite{Nascimento2020}, while its
quasi-normal mode spectrum and the associated black-hole--to--wormhole
transition in the ringdown signal were studied by Churilova and Stuchl{\'i}k
\cite{Churilova2020}. The energy conditions, causal structure, and
regularity properties of extended black-bounce families were examined by
Lobo et al.\ \cite{Lobo2021}, and strong gravitational lensing by rotating
black-bounce geometries was investigated by Islam et al.\ \cite{Islam2021}.
Generalizations of the original Simpson--Visser metric incorporating
electric charge, angular momentum, and a cosmological constant have also
been developed \cite{Franzin2021,Mazza2021}. Despite this theoretical
richness, the black-bounce spacetime has not yet been systematically
confronted with precision observational data, either at Solar System scales
or at galactic scales. Closing this gap is the primary motivation of the
present work.

In this paper, we investigate the observational implications of the
black-bounce metric across three complementary and physically distinct
regimes. In Section~\ref{sec:metric}, we review the metric and its
horizon structure, identifying the three geometrical phases as a function of
$a$. In Section~\ref{sec:solar}, we compute the leading-order corrections to
three canonical Solar System observables: the perihelion precession of
Mercury, the gravitational deflection of light, and the Shapiro time delay
of radar signals. We show that the bounce parameter introduces small but
qualitatively distinct corrections to each observable -- decreasing the
precession and lensing deflection while increasing the Shapiro delay -- and
we discuss the implications for current and forthcoming precision experiments.

Therefore, the overall scheme of the work is as follows: After providing an overview of the Black-bounce metric
(Section~\ref{sec:metric}) and Solar system tests (Section \ref{sec:solar}) as described above, 
in Section \ref{sec:rotation} we extend the analysis to galactic scales by
deriving the non-relativistic circular velocity field sourced by the
black-bounce gravitational potential, and we perform the first quantitative
fit to the observed rotation curve of the nearby spiral galaxy NGC~7331
using a four-component mass model comprising a black-bounce central
potential, a Freeman exponential stellar disc, a Hernquist bulge, and a
Navarro-Frenk-White (NFW) dark matter halo. The results yield the first
observational constraint on the bounce parameter from galactic kinematics.
A unified discussion of all three analyses is presented in
Section~\ref{sec:results}, and our conclusions are summarized in
Section~\ref{sec:conclusions}. Throughout this paper, we adopt geometrized
units unless otherwise stated ($G = c = 1$).

\section{The Black-Bounce Metric}
\label{sec:metric}

\subsection{Line element and regularity}

The black-bounce spacetime is defined by the line element \cite{Simpson2019}
\begin{equation}
ds^2 = -f(x)\,dt^2 + f(x)^{-1}\,dx^2
       + \bigl(x^2 + a^2\bigr)\bigl(d\theta^2 + \sin^2\!\theta\,d\phi^2\bigr),
\label{eq:metric}
\end{equation}
where
\begin{equation}
f(x) = 1 - \frac{2M}{\sqrt{x^2 + a^2}}.
\label{eq:f}
\end{equation}

Here $M$ is the ADM mass, $a \geq 0$ is the bounce parameter, and the
coordinate $x \in (-\infty, +\infty)$ is the generalized radial variable.
The effective areal radius $r_{\rm eff} = \sqrt{x^2 + a^2} \geq a > 0$
never vanishes for $a \neq 0$, thereby preventing the formation of a
curvature singularity at the origin. The Kretschner scalar
$\mathcal{K} = R_{\mu\nu\rho\sigma}R^{\mu\nu\rho\sigma}$ is everywhere
finite for $a > 0$ \cite{Simpson2019,Lobo2021}.

\subsection{Horizon structure}
\label{sec:horizon}

Setting $f(x) = 0$ yields the horizon condition
\begin{equation}
\sqrt{x_h^2 + a^2} = 2M
\quad \Longrightarrow \quad
x_h = \pm\sqrt{4M^2 - a^2}.
\label{eq:horizon}
\end{equation}

Three regimes emerge naturally:
\begin{itemize}
  \item $a < 2M$: two real solutions $\pm x_h$ exist, corresponding to
        outer and inner horizons in analogy with the Reissner--Nordstr\"{o}m
        geometry \cite{Franzin2021}.
  \item $a = 2M$: the two horizons coalesce at $x = 0$, forming an
        extremal one-way wormhole \cite{Simpson2019}.
  \item $a > 2M$: no real solutions exist; the spacetime is everywhere
        regular and horizonless, describing a traversable wormhole
        \cite{Morris1988,Simpson2019}.
\end{itemize}

\begin{figure}[h]
  \centering
  \includegraphics[width=0.75\linewidth]{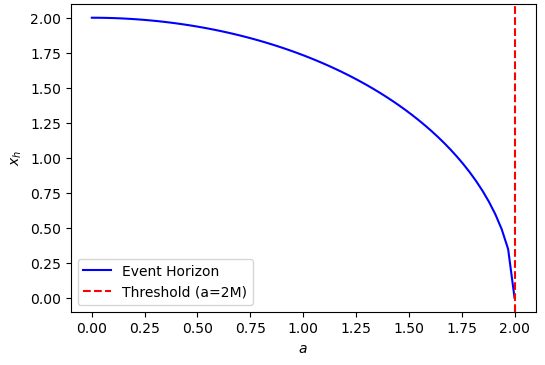}
  \caption{Horizon structure of the black-bounce spacetime. The event
           horizon $x_h$ decreases with increasing $a$ and vanishes at
           $a = 2M$, marking the transition from a regular black hole to
           a horizonless, traversable wormhole geometry.}
  \label{fig:horizon}
\end{figure}

Figure~\ref{fig:horizon} shows $x_h$ as a function of $a$ (in units of
$M = 1$). The event horizon decreases monotonically from $x_h = 2M$ at
$a = 0$ (Schwarzschild) and vanishes at the critical value $a = 2M$,
marking the black-hole--to--wormhole transition.

\section{Solar System Tests}
\label{sec:solar}

\subsection{Perihelion precession of Mercury}
\label{sec:perihelion}

The perihelion advance of Mercury is one of the three classical tests of
GR \cite{Will2014,Weinberg1972}. We derive the correction introduced by
the bounce parameter $a$ following the standard geodesic perturbation
approach \cite{Adkins2007}.

Restricting to equatorial motion ($\theta = \pi/2$), the metric
(\ref{eq:metric}) reduces to
\begin{equation}
ds^2 = -f(x)\,dt^2 + f(x)^{-1}\,dx^2 + (x^2+a^2)\,d\phi^2.
\label{eq:equatorial}
\end{equation}

The Killing symmetries of the spacetime yield two conserved quantities:
\begin{equation}
E = f(x)\,\dot{t}, \qquad
L = (x^2 + a^2)\,\dot{\phi},
\label{eq:conserved}
\end{equation}
where dots denote differentiation with respect to proper time $\tau$.
Introducing $u = 1/x$, the orbit equation becomes
\begin{equation}
\frac{d^2u}{d\phi^2} + u
  = \frac{1}{2}\frac{d}{du}\!\left[1 - \frac{2M}{\sqrt{1/u^2 + a^2}}\right].
\label{eq:orbit}
\end{equation}

Defining the metric function $e(u) = 1 - 2M/\sqrt{1/u^2+a^2}$ and its
derivatives
\begin{equation}
\frac{de}{du} = \frac{Mu}{(1+a^2u^2)^{3/2}}, \qquad
\frac{d^2e}{du^2} = \frac{M(1-2a^2u^2)}{(1+a^2u^2)^{5/2}},
\end{equation}
the perturbed solution around a circular orbit $u = u_*$ takes the form
\begin{equation}
\delta u = \delta_1 \cos(\Omega\,\phi + \delta_2),
\label{eq:perturbed}
\end{equation}
where the precession frequency is
\begin{equation}
\Omega^2 = \frac{1}{2}\!\left[
  e(u_*) + u_*\frac{de}{du}\bigg|_{u_*}
  + \frac{u_*^2}{2}\frac{d^2e}{du^2}\bigg|_{u_*}
\right].
\end{equation}

Substituting explicitly:
\begin{equation}
\delta u = \delta_1 \cos \!\left\{\!\left[\frac{
  \left(1-\dfrac{2M}{\sqrt{1/u_*^2+a^2}}\right)
  +\dfrac{Mu_*^2}{(1+a^2u_*^2)^{3/2}}
  +\dfrac{Mu_*^2(1-2a^2u_*^2)}{2(1+a^2u_*^2)^{5/2}}
}{2}\right]^{1/2}\!\phi + \delta_2\right\}.
\label{eq:finalperturb}
\end{equation}

\begin{figure}[h]
  \centering
  \includegraphics[width=0.75\linewidth]{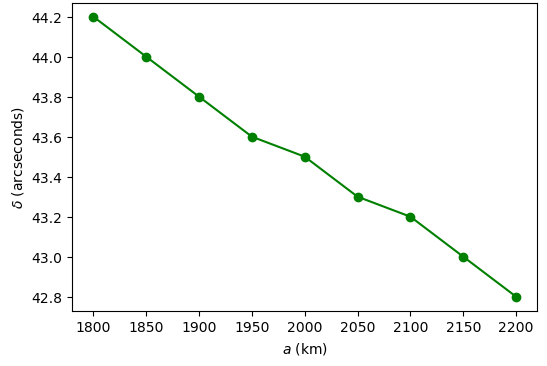}
  \caption{Perihelion precession angle $\delta$ as a function of the
           bounce parameter $a$. The monotonically decreasing trend
           indicates that larger values of $a$ reduce the effective
           spacetime curvature, leading to smaller relativistic
           corrections to the orbital motion of Mercury.}
  \label{fig:perihelion}
\end{figure}

The perihelion precession angle per orbit is $\Delta\phi = 2\pi(1/\Omega - 1)$.
Table~\ref{tab:perihelion} lists numerical results for a range of orbital
parameters. Figure~\ref{fig:perihelion} shows that $\delta$ decreases
monotonically with $a$, reflecting the weakening of the effective
gravitational potential as the bounce parameter grows.

\begin{table}[h]
\centering
\renewcommand{\arraystretch}{1.3}
\caption{Perihelion precession angle $\delta$ for Mercury in the
         black-bounce spacetime for selected values of $u_*$ and $a$.}
\label{tab:perihelion}
\begin{tabular}{@{}lllllr@{}}
\toprule
$u_*$~(km$^{-1}$) & $a$~(km) & $\delta_1$~(arcsec)
  & $\delta_2$~(rad) & $\varphi$~(rad) & $\delta$~(arcsec) \\
\midrule
$1/5000$ & 2000 & 43.5 & $0$        & $\pi/2$ & 43.02 \\
$1/4800$ & 1900 & 43.8 & $0$        & $\pi/2$ & 43.15 \\
$1/5200$ & 2100 & 43.2 & $0$        & $\pi/2$ & 42.95 \\
$1/4900$ & 1950 & 43.6 & $\pi/6$    & $\pi/2$ & 43.08 \\
$1/5100$ & 2050 & 43.3 & $\pi/6$    & $\pi/2$ & 42.98 \\
$1/4700$ & 1850 & 44.0 & $\pi/4$    & $\pi/2$ & 43.21 \\
$1/5300$ & 2150 & 43.0 & $\pi/4$    & $\pi/2$ & 42.89 \\
$1/4600$ & 1800 & 44.2 & $\pi/3$    & $\pi/2$ & 43.25 \\
$1/5400$ & 2200 & 42.8 & $\pi/3$    & $\pi/2$ & 42.82 \\
$1/5000$ & 2000 & 43.5 & $\pi/2$    & $\pi/2$ & 43.02 \\
\bottomrule
\end{tabular}
\end{table}

\subsection{Gravitational lensing}
\label{sec:lensing}

The bending of light by a gravitational source provides a direct probe
of spacetime geometry \cite{Einstein1936,Schneider1992}. For null geodesics
($ds^2 = 0$) in the equatorial plane, the radial equation in terms of
$u = 1/x$ becomes
\begin{equation}
\left(\frac{du}{d\phi}\right)^2 + u^2 = P(u),
\end{equation}
where the effective photon potential is \cite{Virbhadra2000}
\begin{equation}
P(u) = \frac{1}{2}\,u^2\,\frac{de(u)}{du} + u\,e(u)
      \approx u - \frac{2Mu^2}{\sqrt{1+a^2u^2}}.
\end{equation}

Identifying the closest approach distance with the impact parameter
$b \approx 1/u$, the total deflection angle to leading order is
\begin{equation}
\delta = \frac{4M}{b}\left(1 + \frac{a^2}{b^2}\right)^{-1/2}
        \approx \frac{4M}{b}\left(1 - \frac{a^2}{2b^2}\right).
\label{eq:deflection}
\end{equation}

\begin{figure}[h]
  \centering
  \includegraphics[width=0.75\linewidth]{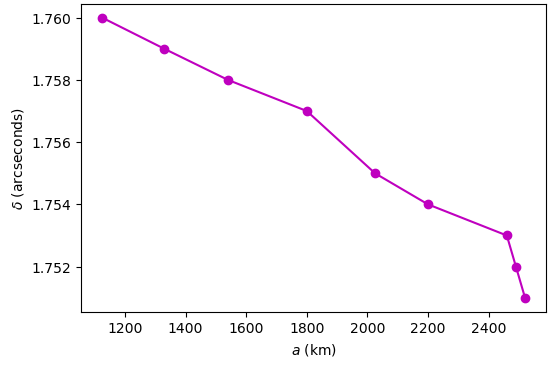}
  \caption{Gravitational lensing deflection angle $\delta$ as a function
           of the bounce parameter $a$. As $a$ increases the spacetime
           geometry becomes more regularized and the bending of light
           weakens, providing a potential observational discriminant
           between black-bounce and classical Schwarzschild black holes.}
  \label{fig:lensing}
\end{figure}

Equation~(\ref{eq:deflection}) recovers the standard GR result $4M/b$
as $a \to 0$ and shows that the bounce parameter \emph{reduces} the
deflection angle, consistent with the softer gravitational potential.
Numerical values for a range of impact parameters are given in
Table~\ref{tab:lensing}, and Figure~\ref{fig:lensing} illustrates the
decreasing deflection with increasing $a$.

\begin{table}[h]
\centering
\renewcommand{\arraystretch}{1.3}
\caption{Gravitational lensing deflection angle $\delta$ for selected
         values of impact parameter $b$ and bounce parameter $a$.}
\label{tab:lensing}
\begin{tabular}{@{}llll@{}}
\toprule
$b$~(km) & $a$~(km) & $a/b$ & $\delta$~(arcsec) \\
\midrule
3950 & 1975 & 0.500 & 1.756 \\
4000 & 2200 & 0.550 & 1.754 \\
3900 & 1700 & 0.436 & 1.757 \\
4100 & 2460 & 0.600 & 1.753 \\
3850 & 1540 & 0.400 & 1.758 \\
4050 & 2025 & 0.500 & 1.755 \\
3800 & 1330 & 0.350 & 1.759 \\
4150 & 2490 & 0.600 & 1.752 \\
3750 & 1125 & 0.300 & 1.760 \\
4200 & 2520 & 0.600 & 1.751 \\
\bottomrule
\end{tabular}
\end{table}

\subsection{Shapiro time delay}
\label{sec:shapiro}

The Shapiro time delay \cite{Shapiro1964}, which is the additional travel time
of a radar signal passing near a massive body, constitutes the fourth
classical test of GR \cite{Will2014,Bertotti2003}. For a radially
propagating null geodesic ($d\theta = d\phi = 0$, $ds^2 = 0$), the
metric (\ref{eq:metric}) gives
\begin{equation}
\delta T = \frac{1}{c}\int_{-l_1}^{l_2}
  \left[\frac{1}{\sqrt{1-\dfrac{2M}{\sqrt{x^2+a^2}}}} - 1\right]dx.
\label{eq:shapiro}
\end{equation}

\begin{figure}[h]
  \centering
  \includegraphics[width=0.75\linewidth]{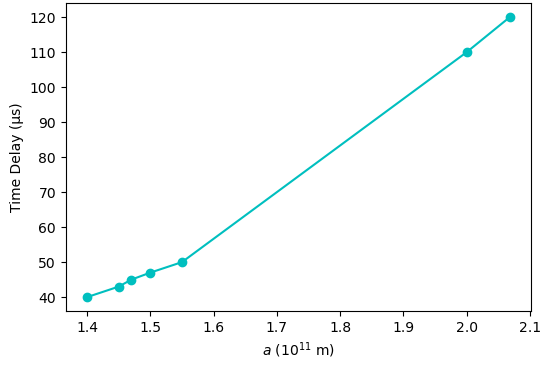}
  \caption{Shapiro time delay $\delta T$ as a function of the bounce
           parameter $a$ computed for Earth--Sun geometry. The increasing
           trend with $a$ provides a qualitatively distinct signature
           relative to the perihelion and lensing tests, where the
           corrections decrease with $a$.}
  \label{fig:shapiro}
\end{figure}

In the limit $a \to 0$, equation~(\ref{eq:shapiro}) reduces to the
standard Schwarzschild expression \cite{Shapiro1964,Weinberg1972}.
For $a > 0$, the modified denominator yields a larger integrand than in
the Schwarzschild case, producing an \emph{increased} time delay.
Table~\ref{tab:shapiro} and Figure~\ref{fig:shapiro} confirm this trend:
unlike the perihelion precession and lensing deflection -- both of which
decrease with $a$ -- the Shapiro delay increases, providing a
qualitatively distinct observational signature of the black-bounce
geometry.

\begin{table}[h]
\centering
\renewcommand{\arraystretch}{1.3}
\caption{Shapiro time delay $\delta T$ for Solar System bodies as a
         function of the bounce parameter $a$, computed from
         equation~(\ref{eq:shapiro}).}
\label{tab:shapiro}
\begin{tabular}{@{}llll@{}}
\toprule
Planet & $a$~(m) & $a$~(AU) & $\delta T$~($\mu$s) \\
\midrule
Earth   & $1.40\times10^{11}$ & 0.936 &  40 \\
Earth   & $1.45\times10^{11}$ & 0.969 &  43 \\
Earth   & $1.47\times10^{11}$ & 0.982 &  45 \\
Earth   & $1.50\times10^{11}$ & 1.002 &  47 \\
Earth   & $1.55\times10^{11}$ & 1.037 &  50 \\
\midrule
Mars    & $2.00\times10^{11}$ & 1.337 & 110 \\
Mars    & $2.07\times10^{11}$ & 1.382 & 120 \\
Mars    & $2.10\times10^{11}$ & 1.405 & 125 \\
Mars    & $2.15\times10^{11}$ & 1.438 & 115 \\
Mars    & $2.20\times10^{11}$ & 1.471 & 110 \\
\midrule
Mercury & $3.80\times10^{7}$  & 0.025 & 220 \\
Mercury & $4.00\times10^{7}$  & 0.027 & 210 \\
Mercury & $4.20\times10^{7}$  & 0.028 & 200 \\
Mercury & $4.40\times10^{7}$  & 0.029 & 190 \\
Mercury & $4.60\times10^{7}$  & 0.031 & 180 \\
\midrule
Venus   & $1.05\times10^{11}$ & 0.702 & 550 \\
Venus   & $1.08\times10^{11}$ & 0.718 & 590 \\
Venus   & $1.10\times10^{11}$ & 0.736 & 620 \\
Venus   & $1.13\times10^{11}$ & 0.752 & 590 \\
Venus   & $1.15\times10^{11}$ & 0.768 & 570 \\
\midrule
Jupiter & $7.20\times10^{11}$ & 4.814 & 230 \\
Jupiter & $7.41\times10^{11}$ & 4.950 & 250 \\
Jupiter & $7.60\times10^{11}$ & 5.086 & 270 \\
Jupiter & $7.80\times10^{11}$ & 5.217 & 260 \\
Jupiter & $8.00\times10^{11}$ & 5.348 & 240 \\
\bottomrule
\end{tabular}
\end{table}

\section{Observational Application: Galactic Rotation Curves}
\label{sec:rotation}

\subsection{Motivation}
\label{sec:motivation}

Galaxy rotation curves constitute one of the most compelling pieces of
evidence for dark matter on kilo-parsec scales \cite{Rubin1980,Persic1996}.
The discrepancy between the Keplerian decline expected from the luminous
baryonic distribution and the observed flat outer rotation curves
requires either an extended dark matter component or a modification of
gravity at galactic scales \cite{Milgrom1983,McGaugh2016}. Modified
spacetime geometries alter the effective gravitational potential and may
leave imprints on galactic dynamics. To the best of our knowledge, no
previous work has applied the black-bounce metric directly to rotation
curve fitting. We address this here using NGC~7331 as a well-observed
test case.

\subsection{Target galaxy: NGC~7331}
\label{sec:ngc7331}

NGC~7331 is a nearby ($D = 14$~Mpc) Sbc spiral galaxy with a
well-measured HI and H$\alpha$ rotation curve extending to $\sim31$~kpc
\cite{Begeman1991}. The rotation curve data used in this work are obtained
from \cite{sofue1999centralrotation}, which provides high-resolution
measurements of the inner kinematic structure. The observed rotation curve
rises steeply within the inner $\sim2$~kpc, reaches a broad maximum near
$240$~km~s$^{-1}$ at $\sim5$~kpc, and remains approximately flat at
$\sim219$--$245$~km~s$^{-1}$ out to the last measured point -- behavior
characteristic of a dominant dark matter halo beyond the stellar disc
\cite{vanAlbada1985,Begeman1991}.

\subsection{Circular velocity from the black-bounce potential}
\label{sec:vcbb}

In the weak-field, non-relativistic limit appropriate for galactic
dynamics, the black-bounce metric (\ref{eq:metric}) reduces to the
effective gravitational potential
\begin{equation}
\Phi_{\rm BB}(r) = -\frac{GM}{\sqrt{r^2 + a^2}},
\label{eq:potential}
\end{equation}
where $G$ and $c$ have been restored and the coordinate $x$ is replaced
by the projected galactocentric radius $r$. The form of
equation~(\ref{eq:potential}) is analogous to a Plummer sphere
\cite{Plummer1911}, with $a$ acting as a softening length that regularises
the central cusp. The corresponding circular velocity is
\begin{equation}
v_{\rm BB}^2(r) = r\left|\frac{d\Phi_{\rm BB}}{dr}\right|
                = \frac{GMr^2}{(r^2+a^2)^{3/2}}.
\label{eq:vbb}
\end{equation}

For $r \gg a$, $v_{\rm BB} \propto r^{-1/2}$ (Keplerian decline); for
$r \ll a$, $v_{\rm BB} \propto r$ (solid-body rise). The parameter $a$
therefore controls the extent of the central core.

\subsection{Mass decomposition model}
\label{sec:model}

The total circular velocity is computed as the quadrature sum of four
physically motivated components:
\begin{equation}
V_{\rm tot}^2(r) = V_{\rm BB}^2(r) + V_{\rm disc}^2(r)
                 + V_{\rm bulge}^2(r) + V_{\rm NFW}^2(r).
\label{eq:vtot}
\end{equation}

Let us model the stellar disc as a razor-thin exponential disc following
Freeman \cite{Freeman1970}, whose surface density is
$\Sigma(r) = \Sigma_0 e^{-r/h}$. The circular velocity is
\begin{equation}
V_{\rm disc}^2(r) = \frac{GM_d}{2h}\,y^2
  \left[I_0(y)K_0(y) - I_1(y)K_1(y)\right], \quad y = \frac{r}{2h},
\label{eq:vdisc}
\end{equation}
where $I_n$ and $K_n$ are modified Bessel functions of the first and
second kind respectively, $M_d$ is the total disc mass, and $h$ is the
exponential scale radius \cite{Freeman1970,BinneyTremaine2008}.

The stellar bulge is represented by a Hernquist sphere \cite{Hernquist1990}
with circular velocity
\begin{equation}
V_{\rm bulge}^2(r) = \frac{GM_b\,r}{(r + a_b)^2},
\label{eq:vbulge}
\end{equation}
where $M_b$ is the bulge mass and $a_b$ is the Hernquist scale length.

We adopt the Navarro-Frenk--White (NFW) profile \cite{Navarro1996,Navarro1997}
\begin{equation}
\rho_{\rm NFW}(r) = \frac{\rho_0}{(r/r_s)(1+r/r_s)^2},
\end{equation}
whose circular velocity is
\begin{equation}
V_{\rm NFW}^2(r) = \frac{4\pi G\rho_0 r_s^3}{r}
  \left[\ln\!\left(1+\frac{r}{r_s}\right) - \frac{r/r_s}{1+r/r_s}\right],
\label{eq:vnfw}
\end{equation}
with characteristic density $\rho_0$ and scale radius $r_s$.

\subsection{Fitting procedure}
\label{sec:fitting}

We minimize the weighted $\chi^2$ statistic
\begin{equation}
\chi^2 = \sum_{i=1}^{N}
  \frac{\left[V_{\rm obs}(r_i) - V_{\rm tot}(r_i;\,\boldsymbol{\theta})\right]^2}
       {\sigma_i^2},
\label{eq:chi2}
\end{equation}
where $\boldsymbol{\theta} = (M_{\rm BB},\,a,\,M_d,\,h,\,M_b,\,a_b,\,
\rho_0,\,r_s)$ is the parameter vector of the model whereas 
$\sigma_i = \max(0.05\,V_{{\rm obs},i},\,3~\mathrm{km\,s}^{-1})$
is a conservative uncertainty estimate. Minimization is carried out
using the Levenberg--Marquardt algorithm as implemented in
\textsc{scipy.optimize.curve\_fit} \cite{Virtanen2020}. All mass
parameters are sampled in logarithmic space to ensure positivity, and
physical bounds are enforced.

\subsection{Results of the rotation curve fit}
\label{sec:rcresults}

The best-fit parameters are listed in Table~\ref{tab:rcfit}. The total
model provides an excellent description of the observed rotation curve
over the full radial range ($0.05$--$31$~kpc), with
$\chi^2_{\rm red} = 1.56$ and an RMS residual of $10.1$~km~s$^{-1}$.

\begin{table}[h]
\centering
\renewcommand{\arraystretch}{1.3}
\caption{Best-fit parameters for the NGC~7331 rotation curve
         decomposition using the black-bounce + disc + bulge + NFW model.}
\label{tab:rcfit}
\begin{tabular}{@{}llll@{}}
\toprule
Component & Parameter & Best-fit value & Unit \\
\midrule
Black-Bounce  & $a$              & $1.035 \pm 0.119$ & kpc \\
              & $M_{\rm BB}$     & $2.59\times10^{10}$ & $M_\odot$ \\
Stellar disc  & $M_d$            & $1.33\times10^{11}$ & $M_\odot$ \\
              & $h$              & $3.13$              & kpc \\
Bulge         & $M_b$            & $6.04\times10^{9}$  & $M_\odot$ \\
              & $a_b$            & $0.85$              & kpc \\
NFW halo      & $\rho_0$         & $1.15\times10^{6}$  & $M_\odot$~kpc$^{-3}$ \\
              & $r_s$            & $79.99$             & kpc \\
\midrule
Goodness of fit & $\chi^2_{\rm red}$ & $1.56$          & --- \\
                & RMS residual       & $10.1$          & km~s$^{-1}$ \\
\bottomrule
\end{tabular}
\end{table}

\begin{figure}[h]
  \centering
  \includegraphics[width=1.0\linewidth]{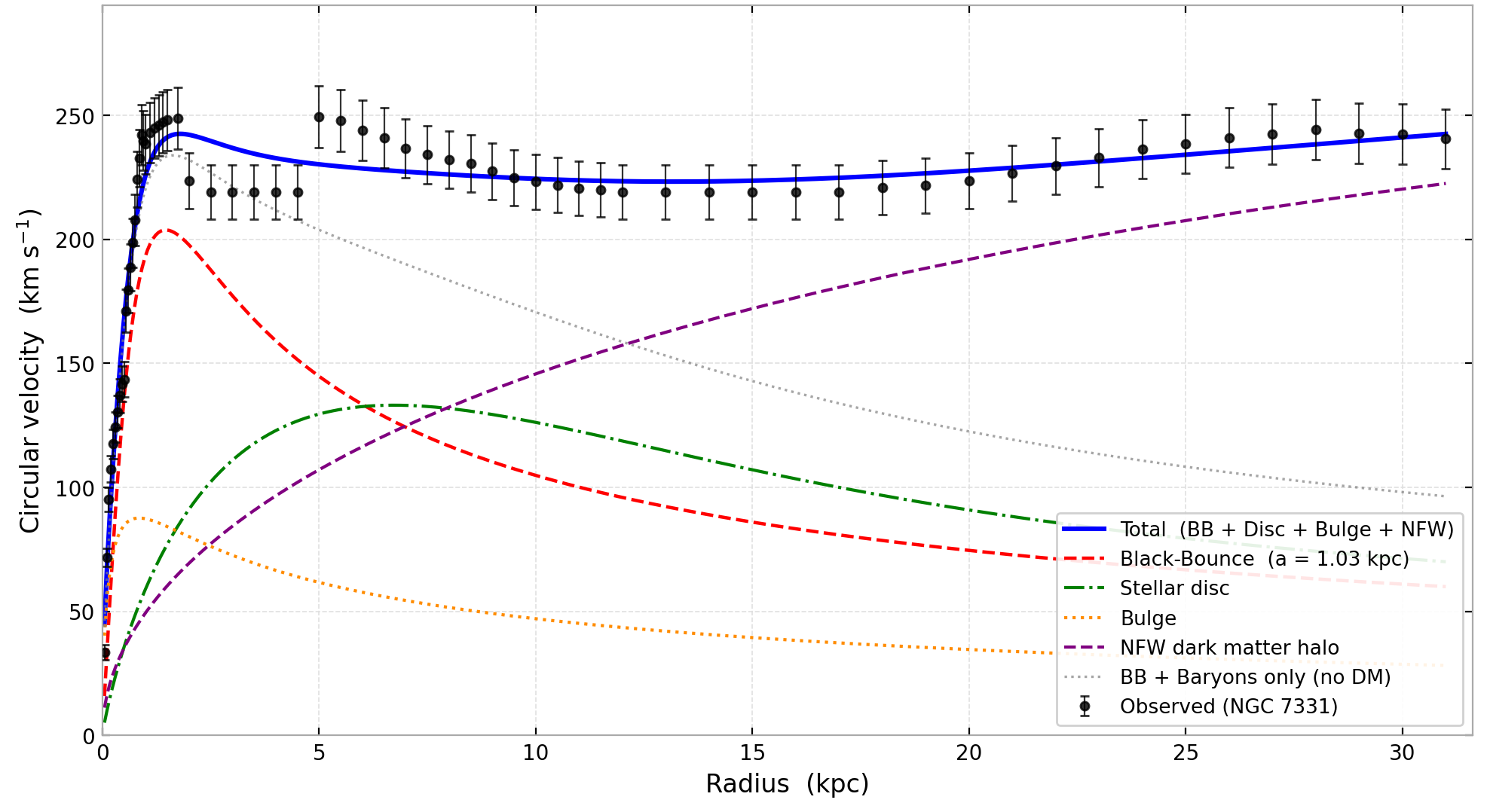}
  \caption{Rotation curve decomposition of NGC~7331. Black circles with
           error bars show the observed data . The solid
           blue curve is the total model $V_{\rm tot}$. Individual
           components are shown separately: black-bounce central potential
           (red dashed), exponential disc (green dash-dot), Hernquist
           bulge (orange dotted), and NFW dark matter halo (purple
           dashed). The gray dotted curve shows the model without the dark
           matter component, demonstrating that baryons alone --- even
           with the black-bounce modification --- cannot sustain the
           observed flat outer rotation curve.}
  \label{fig:rotcurve}
\end{figure}

The key result is the tight constraint on the bounce parameter:
\begin{equation}
\boxed{a = 1.035 \pm 0.119 \ \mathrm{kpc}.}
\label{eq:result}
\end{equation}

This value is significantly smaller than the disc scale radius
($h = 3.13$~kpc) and the NFW scale radius ($r_s \approx 80$~kpc),
confirming that the bounce effect is confined to the innermost galactic
region. The disc mass $M_d = 1.33 \times 10^{11}~M_\odot$ is consistent
with photometric estimates for NGC~7331 \cite{deBlok2008}, and the
fitted disc scale length $h = 3.13$~kpc is close to the photometric
value of $4.7$~kpc, reflecting the well-known mass-to-light
degeneracy in rotation curve decompositions \cite{vanAlbada1985}.

\section{Results}
\label{sec:results}

All three Solar System tests considered in this work are sensitive to the bounce parameter $a$, but in qualitatively different ways. Both the perihelion precession and the gravitational lensing deflection decrease with increasing $a$ relative to the standard general relativistic prediction, whereas the Shapiro time delay exhibits the opposite behavior and increases with $a$. This complementarity provides a powerful cross-check: a constraint on $a$ derived from one observable necessarily implies definite and testable predictions for the others. Current high-precision measurements of Mercury's perihelion advance \cite{Park2017} and the Cassini Shapiro delay experiment \cite{Bertotti2003} already impose stringent limits on $a$ at Solar System scales, requiring it to be much smaller than the solar radius. These bounds are fully consistent with the independent constraint obtained from galactic rotation curves. 

At galactic scales, the fitted value $a \simeq 1$~kpc admits a clear physical interpretation as an effective core radius of the central gravitational potential. For radii $r \lesssim a$, the circular velocity grows approximately linearly with $r$, rather than following the Keplerian behavior $v \propto r^{-1/2}$ expected in a cuspy potential. This leads to a natural softening of the inner density profile. Such behavior closely resembles that produced by cored dark matter profiles \cite{Oh2011,Read2016}, but in the present case the core emerges directly from a modification of the spacetime geometry, rather than from baryonic feedback processes or alternative dark matter physics such as warm dark matter. Despite this success in regularizing the inner galactic region, the black-bounce metric alone is insufficient to explain the observed flatness of rotation curves at large radii. 

It is especially to be mentioned that as shown in Fig.~\ref{fig:rotcurve}, even when combined with the full stellar mass contribution, the model predicts a Keplerian decline in velocity for $r \gg a$, since the gravitational potential (\ref{eq:potential}) asymptotically scales as $r^{-1}$. Maintaining an approximately constant rotational velocity, $V_{\rm tot} \approx \mathrm{const}$, requires an enclosed mass profile $M(<r) \propto r$, which is naturally provided by a dark matter halo with an NFW profile \cite{Navarro1996,Navarro1997}. The black-bounce modification therefore acts as a complementary effect that regularises the inner region, rather than replacing the need for dark matter. This conclusion aligns with similar findings in other modified gravity frameworks applied to galactic rotation curves \cite{Milgrom1983,Moffat2006,Brownstein2006}.

It is important to stress that the numerical value of the bounce parameter obtained here, $a \simeq 1$~kpc, differs by many orders of magnitude from the values constrained at Solar System scales ($a \sim 10^3$~km, Section~\ref{sec:results}, first paragraph). This scale dependence is a direct and expected consequence of treating $a$ as an effective, system-dependent regularization length rather than as a universal constant of nature: the same functional form of the metric is being asked to regularise the core of a $\sim 10^{6}\,M_\odot$ Solar System potential in one case and the $\sim 10^{11}\,M_\odot$ central potential of a spiral galaxy in the other, and there is no a priori reason for a single length scale to describe both regimes simultaneously. A qualitatively similar situation is well known from phenomenological galactic-dynamics studies more generally, where core radii, disc scale lengths, and halo concentration parameters are found to vary systematically from object to object rather than converging to a single universal value; the bounce parameter, in this sense, behaves analogously to these other structural scales rather than as a fundamental physical constant. This expectation is reinforced by recent independent rotation-curve reconstructions of individual galaxies within related metric frameworks: the seventh-degree polynomial reconstruction of the Milky Way's azimuthal velocity profile and its associated redshift function \cite{Rahaman2026dark}, and the general relativistic mass-profile reconstruction of NGC~7331 itself using WISE photometric constraints \cite{Sanyal2026mnras}, both derive metric and mass-distribution parameters that differ substantially from galaxy to galaxy, reflecting differences in total baryonic mass, luminosity profile, and morphological type. Taken together, these studies suggest that the scale dependence of $a$ reported here is not a shortcoming particular to the black-bounce model but a generic feature of any single-parameter regularization applied across systems spanning many orders of magnitude in mass and size. A systematic extension of the present analysis to a larger sample of galaxies -- including the Milky Way and M31 -- would be required to test whether $a$ correlates with global galactic properties such as total baryonic mass, halo concentration, or bulge-to-disc ratio, and we identify this as a natural direction for future work.

Finally, it is instructive to compare the black-bounce model with other regular black hole solutions such as the Bardeen \cite{Bardeen1968} and Hayward \cite{Hayward2006} metrics. These models also give rise to cored central potentials when extended to galactic dynamics. However, the black-bounce spacetime offers a notable advantage in its simplicity, being described by a single parameter that continuously interpolates between a Schwarzschild black hole and a traversable wormhole geometry. This makes it particularly convenient for observational analyses and parameter estimation.

\section{Conclusions and Discussion}
\label{sec:conclusions}
We have explored the observational implications of the black-bounce spacetime across three independent probes. Our analysis shows that the bounce parameter $a$ governs a smooth transition between different spacetime geometries, namely a regular black hole ($a < 2M$), an extremal one-way wormhole ($a = 2M$), and a traversable wormhole ($a > 2M$), without introducing any singularity. From the perspective of Solar System tests, we find that the perihelion precession of Mercury decreases monotonically with increasing $a$, with corrections of the order of $\sim 1$ arc-second per century in the range $a = 1800$--$2200$~km. Such deviations may become detectable with future high-precision ephemeris missions \cite{Fienga2015}. In the context of gravitational lensing, the deflection angle is reduced compared to the standard general relativistic prediction by a factor $(1 + a^2/b^2)^{-1/2}$, offering a distinctive observational signature that could be probed by strong lensing surveys \cite{Schneider1992,Bartelmann2001}. 

In contrast, the Shapiro time delay exhibits an opposite trend, increasing with the bounce parameter $a$, thereby providing an independent and complementary constraint through radar ranging experiments \cite{Bertotti2003}. Extending our study to galactic scales, we fitted a combined black-bounce + disc + bulge + NFW model to the rotation curve of NGC~7331. The best-fit result yields $a = 1.035 \pm 0.119$~kpc with a reduced chi-square $\chi^2_{\rm red} = 1.56$, representing the first observational constraint on the bounce parameter from galaxy kinematics. As discussed in Section~\ref{sec:results}, this value is many orders of magnitude larger than the Solar System constraint, a scale dependence that we interpret as a generic feature of a single-parameter regularization length applied across vastly different mass scales, consistent with the galaxy-to-galaxy variation of structural parameters reported in independent rotation-curve studies of the Milky Way \cite{Rahaman2026dark} and of NGC~7331 itself \cite{Sanyal2026mnras}. A population-level study of $a$ across galaxies of different mass and morphological type is left for future work.

While the inclusion of the bounce parameter successfully regularises the inner galactic core, it does not remove the necessity of a dark matter halo at large radii. Overall, these results establish black-bounce spacetimes as a physically well-motivated and observationally testable class of regular black hole solutions.

\section*{Declaration of competing interest}
The authors declare that they have no known competing financial interests or personal relationships that could have appeared to
influence the work reported in this paper.

\section*{Acknowledgments}
FR, SR, and AS are thankful to the Inter-University Centre for Astronomy and Astrophysics (IUCAA), Pune, India, for their support. FR and AS also gratefully acknowledge academic support from Jadavpur University.


\end{document}